\documentclass[a4paper,twocolumn,11pt]{quantumarticle}
\pdfoutput=1
\usepackage[utf8]{inputenc}
\usepackage[english]{babel}
\usepackage[T1]{fontenc}
\usepackage{amsmath}
\usepackage{hyperref}

\usepackage{tikz}
\usepackage{lipsum}

\begin{document}

\title{When to Reject a Ground State Preparation Algorithm}

\author{Katerina Gratsea}
\affiliation{Zapata Computing Inc.}
\affiliation{ICFO - Institut de Ci\`{e}ncies Fot\`{o}niques, The Barcelona Institute of Science and Technology, Av. Carl Friedrich Gauss 3, 08860 Castelldefels (Barcelona), Spain}

\author{Chong Sun}
\affiliation{Zapata Computing Inc.}

\author{Peter D. Johnson}
\affiliation{Zapata Computing Inc.}

\maketitle

\begin{abstract}
In recent years substantial research effort has been devoted to quantum algorithms for ground state energy estimation (GSEE) in chemistry and materials.
Given the many heuristic and non-heuristic methods being developed, it is challenging to assess what combination of these will ultimately be used in practice.
One important metric for assessing utility is runtime.
For most GSEE algorithms, the runtime depends on the ground state preparation (GSP) method. 
Towards assessing the utility of various combinations of GSEE and GSP methods,
we asked under which conditions a GSP method should be accepted over a reference method, such as the Hartree-Fock state. 
We introduce a criteria for accepting or rejecting a GSP method for the purposes of GSEE. 
We consider different GSP methods ranging from heuristics to algorithms with provable performance guarantees and perform numerical simulations to benchmark their performance on different chemical systems, starting from small molecules like the hydrogen atom to larger systems like the jellium.
In the future this approach may be used to abandon certain VQE ansatzes and other heursitics. Yet so far our findings do not provide evidence against the use of VQE and more expensive heuristic methods, like the low-depth booster. 
This work sets a foundation from which to further explore the requirements to achieve quantum advantage in quantum chemistry.  
\end{abstract}

\section{Introduction}

Quantum computation promises to unlock new
computational capabilities for certain tasks such as the ground state energy estimation (GSEE) for molecules and materials~\cite{Aspuru_Guzik_2005, Goings_2022, QC_for_QA}. 
However, realizing quantum advantage for the task of GSEE requires improvements in the quantum algorithms that will reduce the resource requirements needed, such as the circuit depth~\cite{low_depth_GSEE_Peter}. This has led to extensive research on the development of algorithms with modest circuit depths~\cite{low_depth_GSEE_Peter, LD_new}.

But the performance of GSEE algorithms also strongly depends on the overlap of the true ground state of the Hamiltonian and the initial state generated by a ground state preparation (GSP) method~\cite{Zhang_2022, LD_tables, evidence_paper}. 
For quantum chemistry applications, the Hartree-Fock (HF) Slater determinant state is widely used for GSP, since the cost (in terms of circuit depth) of implementing it on quantum hardware is insignificant compared to GSEE algorithms~\cite{NC_book} and it provides satisfactory results for many molecules and materials~\cite{Google_overlaps}. However, in some important cases the overlap is relatively small (for example molecules with a bond distance out of equilibrium~\cite{Google_overlaps, generative_ground_states}), which creates a need for methods that can provide a larger initial overlap.

To this end, different quantum GSP algorithms have been developed to provide a higher overlap and improve the performance of GSEE algorithms. These include quantum algorithms that prepare multi-determinant states~\cite{Google_overlaps} and numerous GSP algorithms with a provable performance guarantee~\cite{ref13booster, ref14booster, LD_new}. 
Adiabatic state preparation using digital quantum computing \cite{jansen2007bounds, wiebe2012improved} is another approach to preparing ground states and in some works \cite{wecker2015solving, reiher2017elucidating} has been considered the de facto method for state preparation.
More recent heuristic approaches, such as the variational quantum eigensolver (VQE)~\cite{Kottmann_2022, DISCO_VQE} can also provide an approximation to the ground state, while using relatively little circuit depth. Recent work on low-depth-boosters introduced a method with provable performance guarantees on reliably converting circuit depth into ground state overlap and goes beyond heuristic parametrized quantum circuits~\cite{Wang_2022}. They showed that any function f that satisfies the monotonicity suppresses the high-energy eigenstates of a Hamiltonian H and hence boosts the low-energy states~\cite{Wang_2022}. But the choice of the function f and the function parameters is heuristic. Finally, classical machine learning techniques from generative modeling have been applied to the task of generating approximations to the ground state~\cite{generative_ground_states}.

Even though we have a plethora of methods for ground state preparation, we are still missing a reliable way to benchmark their performance. The notion of "good overlap", as usually referred to, is vague and does not explore a performance to resource cost ratio as a benchmarking tool~\cite{Google_overlaps, evidence_paper}. 
We desire benchmark tools that address
the trade-off between the resource cost and performance improvement of the GSP and GSEE subroutines. 
Such a tool could be used to answer questions like:
Is it worth the high circuit depth cost to use a GSP algorithm that provides almost perfect overlap values? Or is it better to settle for a heuristic method like VQE with smaller ground state overlap even though it increases the runtime of the GSEE subroutine? 

The \textit{efficiency} of a GSP algorithm gives the right tools to understand the appropriate balance of resource cost and performance of GSP and GSEE algorithms. To evaluate the performance of quantum algorithms, recent work proposed a resource efficiency metric as the ratio of the success metric over the resource cost~\cite{Auff_ves_2022}. Instead of defining efficiency metrics, in this work we introduce a criteria to evaluate whether to accept or reject a given GSP algorithm. Specifically, we propose a systematic way to benchmark GSP methods for the problem of GSEE. We use the HF method as a reference and explore under which conditions a GSP method will be accepted over HF. The benchmarking criteria incorporate both the reduction of the total runtime for GSEE and the resource cost of the GSP algorithm (see Fig.1). We perform numerical simulations to showcase how to use the criteria in practice and provide a resource estimation of the maximum allowed depth of a GSP to be acceptable over HF.

The paper is organized as follows: in Sec.~\ref{Sec:no_reps} we introduce the criteria for acceptability of GSP that do not require repetitions, while we incorporate repetitions of GSP in Sec.~\ref{Sec:reps}. In Sec.~\ref{Sec:numerics}, we present numerical simulations on how to use the benchmarking criteria set in the first two sections. Also, we include a resource estimation of the maximum allowed GSP circuit depth for solid-state materials. Sec.~\ref{discussion}, contains the conclusions and future research directions.

\begin{figure*}[ht]
    \centering
    \includegraphics[width=0.7\linewidth]{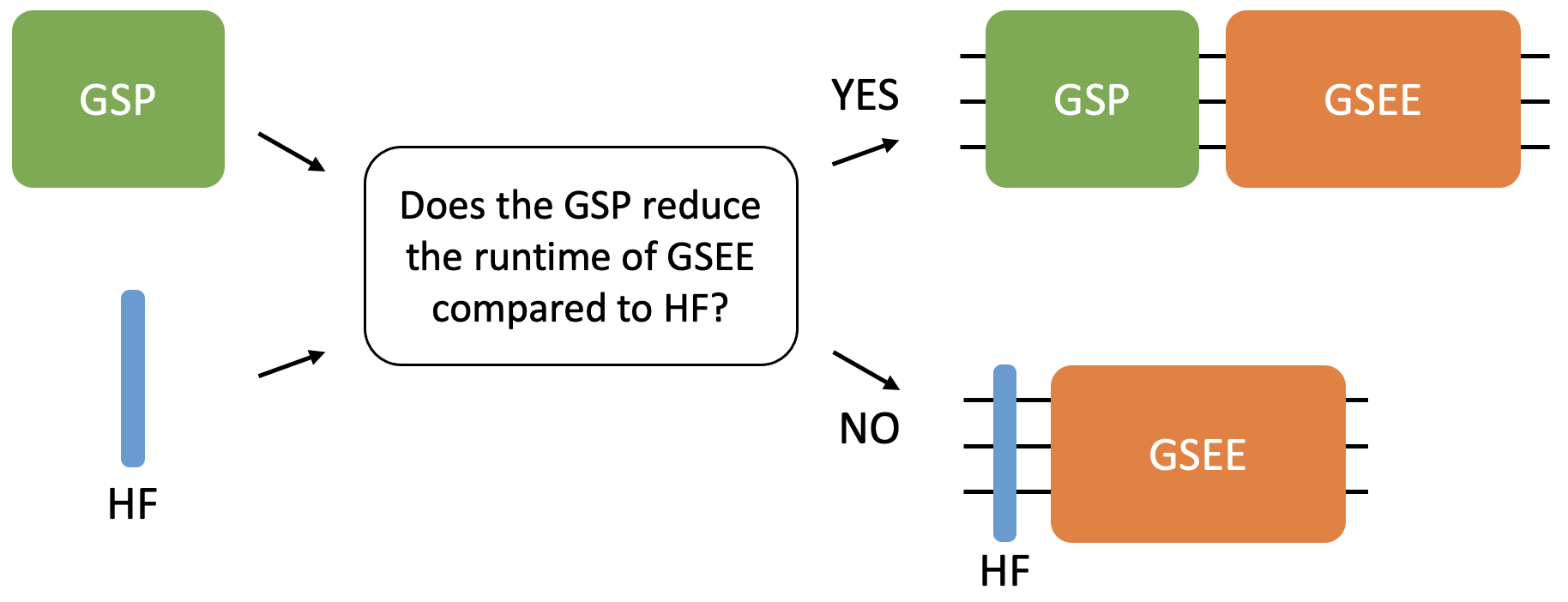}
    \caption{The acceptability criteria is used to benchmark the given GSP method over the HF reference with the goal of reducing the total runtime of the GSEE algorithm.}
    \label{fig:intro}
\end{figure*}

\section{Criteria for acceptability of state preparation}\label{Sec:no_reps}

Here we discuss the criteria under which a state preparation method is acceptable for the purposes of GSEE.
Any ground state energy estimation method has a runtime that depends on the features of the ground state preparation: the GSP circuit depth $D$ and the overlap between the prepared state and the ground state $\eta=\gamma^2$.
The runtime also depends on the target accuracy $\epsilon$. 
For a given energy estimation algorithm~\cite{LD_tables}, the runtime can be formulated as
\begin{align}\label{eq::gsp_no_reps}
    \cal{T}(D,\gamma;\epsilon) &= (\textup{number of repetitions}) \nonumber\\
    \times&(\textup{total circuit depth of each repetition}) \nonumber\\
    &= (\textup{number of repetitions}) \nonumber\\
    \times&(\textup{GSP depth} +\textup{GSEE depth})\nonumber\\
    &=\tilde{O}\left(\frac{1}{\gamma^\alpha}\left(D+\frac{1}{\epsilon\gamma^{\beta}}\right)\right).
\end{align}
According to the table given in reference~\cite{LD_tables}, example GSEE algorithms have values $\alpha \in \{0,2,4 \}$ and $\beta \in \{0, 1,2 \}$.
The units for the GSP and GSEE circuit depths need to match, but as it will become evident from the numerical simulations in the next section different choices for the units could be used, such as the circuit depth and the T-gate count. Also, we are ignoring constant factors and logarithmic dependence on the parameters for now to simplify the introduction of this technique;
but these should ultimately be included to set a more accurate benchmark. Finally, some state preparation methods require a number of repetitions to ensure their success (with high probability). In Sec.~\ref{Sec:reps}, we discuss the runtime cost that includes repetitions of the GSP algorithms.

To establish the concept of an \emph{acceptable} state preparation, we must assume that the ground state energy estimation algorithm has a reference or default method to compare with. 
A reference initial state could be simply a product state of the measurement basis, or the solution to a mean-field-level method, such as the Hartree-Fock ground state.
The circuit depth for preparing a reference state can usually be neglected, thus we label the depth and overlap as $D_0=0$ and $\gamma_0$. In this work, we focus on the HF method as a reference exactly due to the zero depth cost. Other methods, such as adiabatic state preparation~\cite{Aspuru_Guzik_2005, Albash_2018}, which have non-zero depth cost, could be used as reference in future work. 
The runtime of GSEE using the HF reference state preparation is
\begin{align}\label{reference_runtime}
    \mathcal{T}_0(D_0 ,\gamma_0;\epsilon)
    &=\tilde{O}\left(\frac{1}{\epsilon\gamma_0^{\alpha+\beta}}\right).
\end{align}
The condition for a state preparation method to be acceptable over the reference is that the total runtime of the GSEE with the GSP $\left( \mathcal{T} \right)$ is smaller than the total runtime of the GSEE with HF $\left( \mathcal{T}_0 \right)$, i.e. $\mathcal{T}<\mathcal{T}_0$. This puts constraints on the state preparation parameters:
\begin{align}\label{general_condition}
    \frac{1}{\gamma^\alpha}\left(D+\frac{1}{\epsilon\gamma^{\beta}}\right)<\frac{1}{\epsilon\gamma_0^{\alpha+\beta}}.
\end{align}
Observe that if $D=0$, then the acceptability criteria reduce to the condition $\gamma>\gamma_0$.
We can rewrite the general condition~\eqref{general_condition} as
\begin{align}
    D<\frac{1}{\epsilon\gamma^{\beta}}\left(\left(\frac{\gamma}{\gamma_0}\right)^{\alpha+\beta}-1\right).
\end{align}
This shows that if $\epsilon$ is decreased then a state preparation with larger $D$ will be accepted. In other words, for less demanding GSEE algorithms with a worse target accuracy $\epsilon$ more costly GSP algorithm could be accepted over the HF.
Finally, we could write the above inequality as following
\begin{equation} \label{eq:criteria}
    \frac{D+1/\epsilon\gamma^{\beta}}{1/\epsilon\gamma^{\beta}}<\left(\frac{\gamma}{\gamma_0}\right)^{\alpha+\beta}.   
\end{equation}
Therefore, the acceptability criterion in the more strict case when $\alpha + \beta=1$, could be expressed in words as
\begin{align}
\frac{\textup{total depth}}{\textup{GSEE depth}}<\frac{\textup{gsp overlap}}{\textup{HF overlap}}
\end{align}
or
\begin{align}
\frac{\textup{total depth}}{\textup{GSEE depth}}<\frac{\textup{$N_{reps}$ from HF}}{\textup{$N_{0, reps}$ from GSP}}, 
\end{align}
where $N_{reps}$ is the number of repetitions due to the GSP overlap value, i.e. $N_{reps}=1/\gamma$ and $N_{0, reps}=1/\gamma_0$.

Next, we discuss the simple case when the GSP query depth is much smaller than the GSEE query depth. Then, we have
\begin{align}
    \frac{\textup{GSP depth}}{\textup{GSEE depth}} \ll 1 \\
    D \gamma^\beta \ll \frac{1}{\epsilon}.
\end{align}

For typical values of $\epsilon \simeq 10^{-3}$~\cite{low_depth_GSEE_Peter,Babbush_2018}, the condition becomes $D \gamma^\beta \ll 10^3$. Since $\beta \in {\{0,1,2\}}$~\cite{LD_tables} and $\gamma \leq 1$, the more strict condition is $D \ll \dfrac{1}{\epsilon}$.

Then, the acceptability criteria for any $\alpha \in \{0,2,4 \}, \beta \in \{0,1,2\}$~\cite{LD_tables} is simplified to 
\begin{align}\label{eq::simple_acc_criteria}
    1 <\left(\frac{\gamma}{\gamma_0}\right)^{\alpha+\beta},
\end{align}
which simply states that the acceptance of the GSP over HF is determined by the respective overlap values ratio.

Next, we compare the acceptability criteria for two different GSEE algorithms presented in the table of GSEE performance~\cite{LD_tables}, the quantum phase estimation semi-classical (QPE)~\cite{QPE_semclassical1, QPE_semiclassical2} and the GSEE algorithm developed by Tong et. al. (referred as LT20)~\cite{LD_tables} for a given GSP algorithm introduced in~\cite{ref14booster} whose depth depends on the lower bound of the spectral gap $\Delta$ and the overlap $\gamma_0$. For QPE we have $\alpha=\beta=2$ and
\begin{equation}
\frac{ \epsilon  \gamma^{2}+\Delta \gamma_0}{\Delta \gamma_0}<\left(\frac{\gamma}{\gamma_0}\right)^{4},
\end{equation}
while for LT20 we have $\alpha=0 $, $\beta=1$ and
\begin{equation}
\frac{ \epsilon  \gamma+\Delta \gamma_0}{\Delta \gamma_0}<\left(\frac{\gamma}{\gamma_0}\right).
\end{equation}
Since $\gamma \leq 1$, we have
\begin{equation}
\frac{ \epsilon  \gamma^{2}+\Delta \gamma_0}{\Delta \gamma_0} \leq \frac{ \epsilon  \gamma+\Delta \gamma_0}{\Delta \gamma_0} < \left(\frac{\gamma}{\gamma_0}\right) < \left(\frac{\gamma}{\gamma_0}\right)^{4}.
\end{equation}
This suggests that the acceptability criteria for the LT20 is more strict than the QPE. The LT20 has a smaller GSEE query depth compared to QPE, so it is harder to accept a GSP algorithm with non-zero depth over the HF. Therefore, the better the GSEE algorithms becomes in terms of query depth reduction, the more strict the criteria for the acceptance of a GSP method over HF. In other words, as the query depth of the GSEE becomes smaller, the number of repetitions imposed by the overlap prepared from the GSP algorithm becomes less significant.

Finally, the acceptability criteria allow us to explore the maximum values of the GSP depth that enable the given GSP method to be acceptable over the HF state. Given a specific GSEE algorithm and the value of $\gamma_0$, and assuming that the GSP provides a specific value of $\gamma$, i.e. $\gamma=1$, we find the corresponding maximum acceptable depth of a GSP method. To this end, Eq.~\eqref{eq:criteria} can be written as 
\begin{equation}\label{eq:RE}
    D < \frac{\gamma - \gamma_0}{\gamma_0} D_\mathrm{GSEE},
\end{equation}
for more demanding GSEE algorithms with $\alpha + \beta =1$ and depth $D_\mathrm{GSEE}$.
The above equation can be expressed as
\begin{equation}\label{eq::Dgps}
\textup{GSP depth}  < \frac{\textup{performance gain}}{\textup{HF performance}}\,  \textup{GSEE depth }.
\end{equation}

\section{Criteria for acceptability of state preparation methods with repetitions}\label{Sec:reps}

In this section, we discuss the criteria under which a GSP method that requires repetitions to reach an overlap $\gamma$ is acceptable for a given GSEE algorithm.
The runtime Eq.~\eqref{eq::gsp_no_reps} discussed in the previous section becomes
\begin{align}
    \cal{T}(D,\gamma;\epsilon) &= (\textup{number of repetitions of GSEE}) \nonumber\\
    \times & [ (\textup{number of repetitions of GSP}) \nonumber\\
    \times & (\textup{circuit depth of GSP}) \nonumber\\
    + & (\textup{circuit depth of GSEE}) ]\nonumber\\
    &= \tilde{O} \left( \frac{1}{\gamma^\alpha}\left( \frac{1}{P_\mathrm{succ}} \times  D  + \frac{1}{\epsilon\gamma^{\beta}} \right) \right).
\end{align}

As explained earlier, the runtime for the GSEE using the HF reference state preparation $\mathcal{T}_0 $ is given by Eq.~\eqref{reference_runtime}. Then, the condition, i.e. $\mathcal{T}<\mathcal{T}_0$, for a state preparation method being acceptable over the reference becomes

\begin{multline}\label{eq::gsp_with_reps}
\frac{ D/  P_{succ}+1/\epsilon\gamma^{\beta}}{1/\epsilon\gamma^{\beta}}<\left(\frac{\gamma}{\gamma_0}\right)^{\alpha+\beta},
\end{multline}
which can be expressed as
\begin{equation*}
\frac{\textup{total query depth}}{\textup{GSEE query depth}}<\frac{\textup{$N_\mathrm{reps}$ from HF}}{\textup{$N_{reps}$ from GSP}}.
\end{equation*}

\section{Numerical simulations}\label{Sec:numerics}

In this section, we apply the acceptability criteria and benchmark different GSP methods over the HF for different Hamiltonians of molecules and solid-state materials, starting from small molecules ($\textup{H}_2$ molecule) and moving on to larger molecules ($\textup{N}_2$ molecule). Finally, we perform a resource estimation of the maximal acceptable circuit depth of GSP for different solid-state materials over HF state.

\subsection{Molecules}

We explore the acceptability criteria for a small molecule ($\textup{H}_2$) with $4$ spin-orbitals or qubits in an adapted basis~\cite{Kottmann_2021}. We compare the HF method to the separable pair approximation (SPA) approach introduced in the recent work of~\cite{Kottmann_2022} as the GSP method (see Appendix~\ref{appendix}). According to this work, the circuit depth of SPA for the $\textup{H}_2$ molecule equals to 3. Since the depth is $D_{GSP}=3 << 10^3$ for typical values of chemical accuracy $\epsilon= 10^{-3}$, we are in the simple case discussed in Sec.~\ref{Sec:no_reps} where the criteria are simplified to the overlap values ratio (Eq.~\eqref{eq::simple_acc_criteria}). We assume a more demanding GSEE algorithms with $\alpha+\beta=1$ and the criteria is given by 
\begin{align}\label{eq::simple_acc_criteria2}
    1 <\frac{\gamma}{\gamma_0}.
\end{align}

\begin{figure}[ht]
    \centering
    \includegraphics[width=1.0\linewidth]{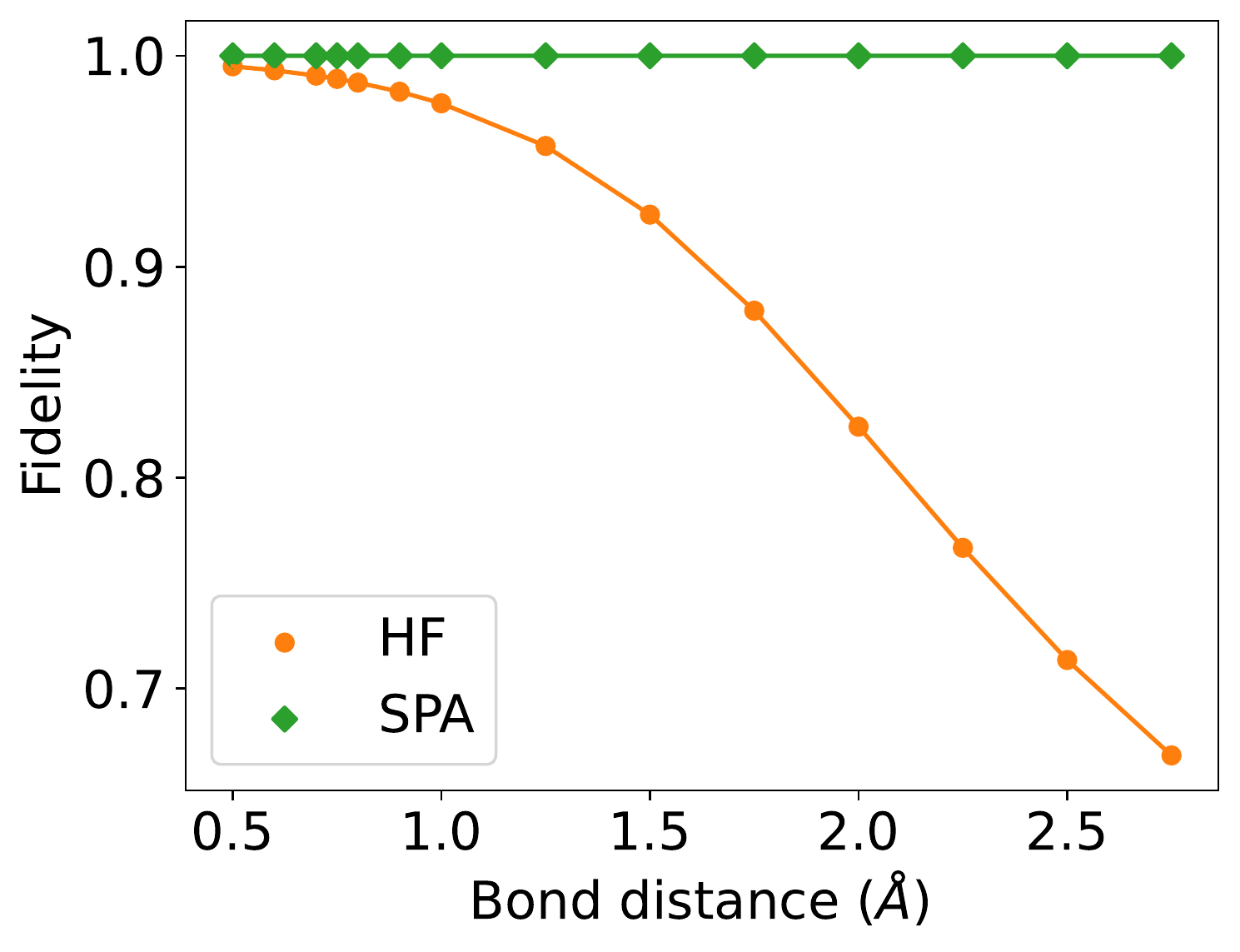}
    \caption{Fidelity as a function of bond distance of the $\textup{H}_2$ molecule for the HF method and the SPA algorithm as GSP methods respectively. The value of alpha and beta is set to be alpha + beta = 1, which corresponds to a GSEE method that has a more strict acceptability criteria.}
    \label{fig:H2_molecule}
\end{figure}

In Fig.\ref{fig:H2_molecule}, we plot the fidelity of the two different GSP methods.  For all bond distances presented in Fig.\ref{fig:H2_molecule}, the criteria is satisfied. Specifically, at bond distance $d=0.5$ the ratio $\frac{\gamma}{\gamma_0}$ is $1.005$ leading up to the value of $1.5$ for $d=2.6$. This suggests that initially SPA is comparable to HF and as we increase the bond distance (\r{A}), SPA is acceptable over HF. For a less demanding GSEE algorithm (i.e. with $\alpha+\beta$ possessing different values than $1$), the criteria would be satisfied and SPA would be acceptable over the HF method.

Next, we explore the larger molecule $N_2$ considering $6$ active electrons in $12$  spin-orbitals or qubits in an adapted basis~\cite{Kottmann_2021} at bond distance $d=2.0$. Initially, we benchmark the SPA method over the HF (see Appendix~\ref{appendix}). Since $D_{GSP}=3$~\cite{Kottmann_2022}, we could apply the criteria of Eq.\ref{eq::simple_acc_criteria2} which gives

\begin{equation}
    1 < \frac{\gamma}{\gamma_0} = \frac{0.85}{0.72} = 1.18.
\end{equation}
Therefore, the SPA is acceptable over the HF method. 

Next, we benchmark a more costly heuristic algorithm: the low-depth-booster from the recent work of~\cite{Wang_2022}. To this end, we change the unit of depth from the circuit depth to the accumulations of the controlled time evolution $epx\left( 2 i \pi H \right)$ operations, where H is the Hamiltonian of the system. Following the aforementioned work~\cite{Wang_2022}, we have $D_{GSP}=10^3$ with $\gamma \approx 1$, while $D_{GSEE}=2 \times 10^4$ and $\gamma_0 = 0.72$. The success probability of the low-depth-booster GSP algorithm applied with the linear combination of unitaries (LCU) method is $P_{succ} \approx 0.5$~\cite{Wang_2022}. Therefore, the criteria of Eq.~\ref{eq::gsp_with_reps} becomes
\begin{align}
    \frac{ D/ \left( P_{succ}\right)+1/\epsilon\gamma^{\beta}}{1/\epsilon\gamma^{\beta}}<\left(\frac{\gamma}{\gamma_0}\right)^{\alpha+\beta} \Rightarrow \\
    \frac{2.2}{2}<\left(\frac{1}{0.72}\right)^{\alpha+\beta} \Rightarrow
    1.1<\left(1.39\right)^{\alpha+\beta},
\end{align} 
which is satisfied for any values of $\alpha, \beta$ of the GSEE algorithms.

\subsection{Solid-state materials}

Here, we perform a resource estimation of the maximal acceptable depth of the GSP methods for different solid-state materials.

Recent work~\cite{Babbush_2018} estimates the T gates needed for quantum simulation of 3D spinful jellium (or the homogeneous electron gas). It focuses on T-count since applying a T gate requires a lot of logical qubits and takes much longer than any other operation in a quantum circuit~\cite{T_gates}. The 3D spinful jellium is in the dual basis at Wigner-Seitz radius of 10 Bohr radii assuming the system is at half filling.  For $54$ spin-orbitals and a target chemical accuracy $\Delta E = 0.0016$ Hartree the depth is equal to $1.8\times 10^7$ T-count.

\begin{figure}[ht]
    \centering
    \includegraphics[width=1.0\linewidth]{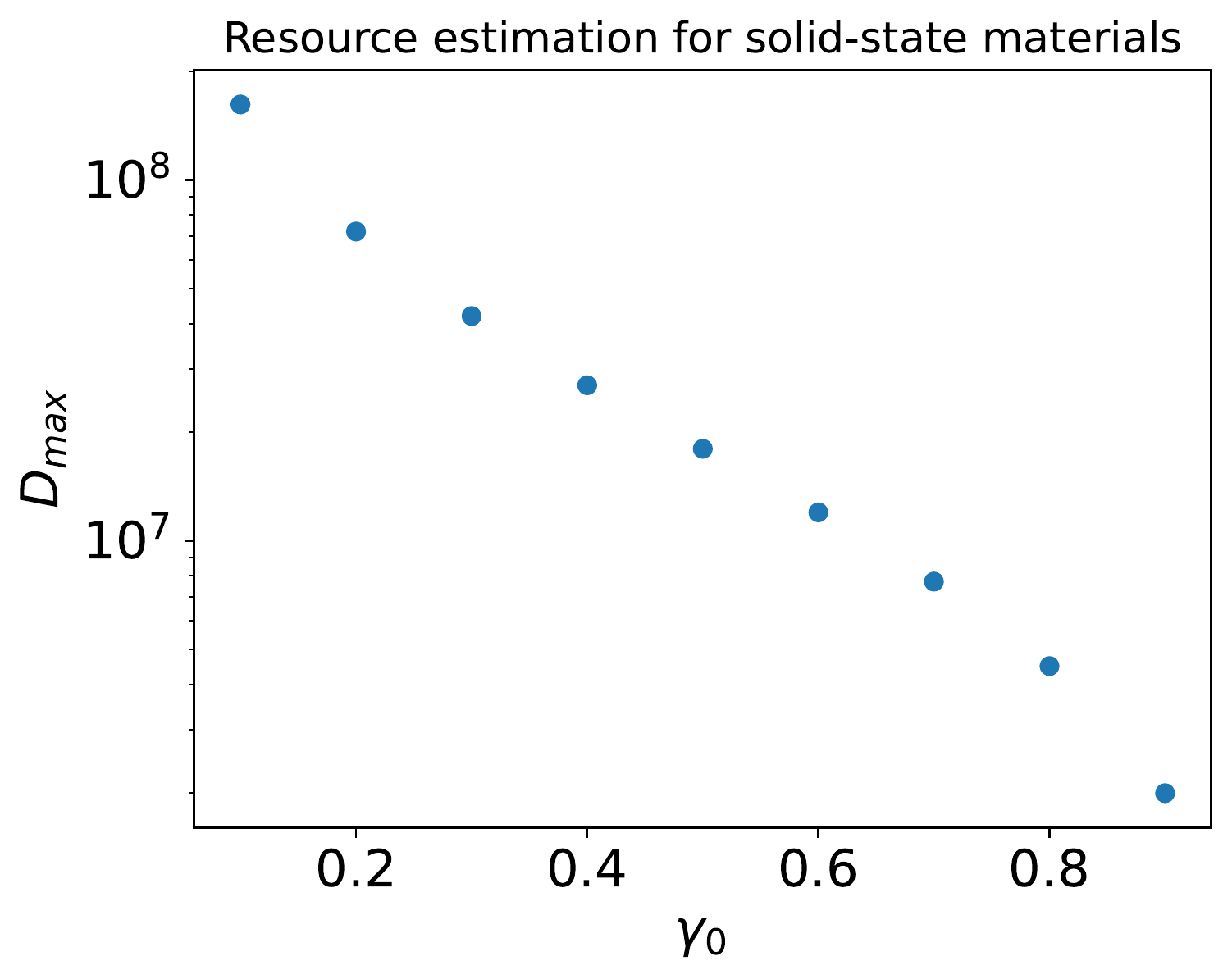}
    \caption{Maximum acceptable depth $D_{max}$ of the GSP algorithm with $\gamma=1$ and $D_{GSEE} = 1.8 \times 10^7$ for solid state materials as a function of the HF overlap $\gamma_0$.}
    \label{fig:jellium}
\end{figure}

As explained in the work of~\cite{Babbush_2018}, the jellium is a good proxy for different solid-state materials, such as diamond, graphite, silicon, metallic lithium and crystalline lithium hydride. For these materials, the HF overlap could range from smaller to larger values as presented in Fig.~\ref{fig:jellium}. Assuming that the GSP method gives $\gamma \approx 1$, we have a resource estimation of the maximum depth allowed for the GSP method to be acceptable over HF (see Fig.~\ref{fig:jellium}) given by Eq.~\eqref{eq:RE}.

\section{Discussion}\label{discussion}

We introduced a method to assess when to accept or reject a ground state preparation (GSP) method over the Hartree-Fock (HF) reference for the task of ground state energy estimation (GSEE) by introducing acceptability criteria. The criteria are defined through the total runtime of the GSEE algorithm that incorporates both the number of repetitions needed and the total circuit depth of each repetition---i.e. the GSP and the GSEE depth. If the inequality introduced in Eq.~\ref{eq:criteria} is satisfied, then the GSP method is acceptable over the HF---i.e. provides a speedup in the total runtime of the GSEE algorithm. The criteria explores the trade-off of both the resource cost and performance of GSP and GSEE subroutines.

We explored under which conditions the acceptability criteria could be simplified and also established them for GSP methods that require repetitions to reach an overlap $\gamma$. Comparing the acceptability criteria for two different GSEE algorithms with a GSP, we found that the better the GSEE algorithms becomes in terms of query depth reduction, the more strict the criteria to accept a GSP over the HF. This could be due to the fact that the number of repetitions introduced by the GSP overlap becomes less significant as the GSEE query depth becomes smaller. The ability to trade circuit depth with runtime is also motivated in recent works~\cite{low_depth_GSEE_Peter, LD_new}. In agreement, the resource estimation performed in this work suggests that a GSP method with a larger circuit depth than the GSEE could be accepted for total runtime reduction.

Next, we showed that the separable pair approximation (SPA) method is acceptable over the HF for the hydrogen molecule for different bond lengths, which suggests that even for a simple molecule, there exists GSP that could offer an improved performance to the GSEE algorithm over using HF. We also evaluated the more expensive low-depth-booster GSP algorithm for the nitrogen molecule which is widely used for benchmarking quantum chemistry simulations; in particular, when the bond is stretched~\cite{Google_overlaps, Kottmann_2022, Wang_2022}. These results suggest that more expensive GSP methods could reduce the total runtime, thus being acceptable over the HF reference. In accordance, the resource estimation of the maximum allowed depth of a GSP does not provide evidence against the use of VQE and more expensive heuristic methods. 
Further numerical and theoretical work is needed to draw a more definitive conclusion.


This work sets 
a foundation
to further explore resource efficiency metrics for GSP and GSEE algorithms. It would be interesting to apply the criteria introduced here to molecules and materials of industrial relevance~\cite{Gonthier_2022} and further use them for resource estimations. Moreover, they could be adjusted to incorporate logarithmic dependencies on the parameters or the recent GSEE algorithm with an exponential improvement in the circuit depth~\cite{low_depth_GSEE_Peter}. Finally, other GSP methods could play the role of the reference method instead of the HF. It is challenging to assess what combination of methods will ultimately be used in practice, and further research will help evaluate the utility of various combinations of GSEE and GSP methods.

\noindent\textbf{Acknowledgements} This work was done while K.G. was a research intern at Zapata Computing Inc. We thank Artur Izmaylov, Jerome Gonthier, and Jakob Kottmann for helpful discussions. K.G. acknowledges support from: European Union’s Horizon 2020 research and innovation programme under the Marie Skłodowska-Curie grant agreement No 847517, Ministerio de Ciencia y Innovation Agencia Estatal de Investigaciones (R$\&$D project CEX2019-000910-S AEI/1013039/5011000011033, Plan National FIDEUA PID2019-106901GB-I00, FPI), Fundació Privada Cellex, Fundació Mir-Puig, and from Generalitat de Catalunya (AGAUR Grant No. 2017 SGR 1341, CERCA program).
\bibliographystyle{unsrt}
\bibliography{references}

\appendix

\section{Computational details}\label{appendix}
For the numerical simulations we followed the \href{https://github.com/tequilahub/tequila-tutorials/blob/main/chemistry/SeparablePairAnsatz.ipynb}{notebook}~\cite{Kottmann_2022} and the molecular \href{https://github.com/tequilahub/tequila-tutorials/tree/main/chemistry/data/basis-set-free-molecules}{data}~\cite{Kottmann_2021}.

\end{document}